# Empirical Evaluation of Embedding Models in the Context of Text Classification in Document Review in Construction Delay Disputes

Fusheng Wei
Data & Technology
Ankura Consulting Group, LLC
Washington, D.C. USA
fusheng.wei@ankura.com

Robert Neary
Data & Technology
Ankura Consulting Group, LLC
Washington, D.C. USA
robert.neary@ankura.com

Han Qin
Data & Technology
Ankura Consulting Group, LLC
Washington, D.C. USA
han.qin@ankura.com

Qiang Mao
Data & Technology
Ankura Consulting Group, LLC
Washington, D.C. USA
qiang.mao@ankura.com

Jianping Zhang
Data & Technology
Ankura Consulting Group, LLC
Washington, D.C. USA
jianping.zhang@ankura.com

*Abstract*— Text embeddings are numerical representations of text data, where words, phrases, or entire documents are converted into vectors of real numbers. These embeddings capture semantic meanings and relationships between text elements in a continuous vector space. The primary goal of text embeddings is to enable the processing of text data by machine learning models, which require numerical input. Numerous embedding models have been developed for various applications. This paper presents our work in evaluating different embeddings through a comprehensive comparative analysis of four distinct models, focusing on their text classification efficacy. We employ both K-Nearest Neighbors (KNN) and Logistic Regression (LR) to perform binary classification tasks, specifically determining whether a text snippet is associated with 'delay' or 'not delay' within a labeled dataset. Our research explores the use of text snippet embeddings for training supervised text classification models to identify delay-related statements during the document review process of construction delay disputes. The results of this study highlight the potential of embedding models to enhance the efficiency and accuracy of document analysis in legal contexts, paving the way for more informed decision-making in complex investigative scenarios.

## I. Introduction

In the realm of complex legal disputes, such as antitrust investigations, legal practitioners face the daunting challenge of analyzing extensive collections of documents to uncover critical evidence and derive actionable insights. The sheer volume and intricate nature of these documents demand sophisticated techniques for efficient and accurate analysis. Large language models (LLMs) have emerged as powerful tools in addressing various natural language processing (NLP) tasks, including text classification. A key aspect of these tasks is the use of text embeddings. Text embeddings are numerical representations of text data, where words, phrases, or entire documents are converted into vectors of real numbers. These embeddings capture semantic meanings and relationships between text elements in a continuous vector space. The primary goal of text embeddings is to enable the processing of text data by machine learning models, which require numerical input.

The landscape of embedding techniques is broad, covering both traditional and modern approaches. Traditional methods, such as Bag of Words and Term Frequency-Inverse Document Frequency (TFIDF), are valued for their simplicity and speed, making them popular choices in many text analysis tasks. However, recent advancements in embeddings derived from LLMs offer a more sophisticated representation of language, often surpassing traditional methods in capturing the subtleties of textual content. These modern embeddings leverage deep neural networks and contextual information to provide a deeper semantic understanding, which is particularly beneficial in complex and nuanced domains like legal analysis.

This study focuses on evaluating three popular embedding methods—bag-of-word, Sentence Transformers, and NVIDIA NeMo's Large Language Model (LLM)-based embeddings—to assess their impact on predictive modeling, within the context of real-world legal datasets. By evaluating the strengths and weaknesses of each approach, this paper seeks to provide a comprehensive assessment of their effectiveness and suitability for legal text classification. Through this comparative analysis, we aim to identify optimal strategies for utilizing embeddings to extract meaningful insights from complex legal documents, thereby enhancing the efficiency and accuracy of legal investigations.

## II. Embedding Models for Text Classification

### A. Bag-of-Word

The bag-of-words approach is a popular method for representing a text message. In this approach, a piece of text message is represented as a vector of feature values—a numerical representation. The text (e.g., sentence or document) is treated as a 'bag' of its words, ignoring grammar and word order while accounting for the frequency of word occurrences.

Each unique word (or phrase), known as a 'token', serves as a feature. This method provides a simplified representation of the text across the document collection.

A traditional statistical method, TF-IDF assigns weights to words (features) based on their frequency in a document relative to their occurrence in the corpus. While simple and computationally efficient, it does not capture semantic relationships between words.

*B. Sentence Transformers*

Leveraging advancements in transformer architectures like BERT and RoBERTa, Sentence Transformers generate contextual embeddings at the sentence level. These embeddings encode semantic meaning, making them effective for downstream tasks such as classification and clustering [1].

*C. NVIDIA NeMo Embeddings*

NVIDIA NeMo offers state-of-the-art LLM embeddings that are fine-tuned on diverse datasets. These embeddings capture rich contextual and semantic nuances, potentially outperforming traditional methods in tasks involving complex language understanding [2] [3][5].

*D. Model comparions*

In this paper, we used four embedding models: Bag-of-Word, two sentence transformer models: all-MiniLM-L6-v2 and nli-mpnet-base-v2 [4], and nv-embed-v2. The bag-of-word model generates term-document matrices and scales efficiently with the size of the dataset, providing a straightforward, non-parametric approach to text representation.

In contrast, all-MiniLM-L6-v2 serves as a lightweight and fast solution for general semantic tasks, delivering reasonable accuracy. It is a powerful tool for sentence and short paragraph encoding, capable of mapping sentences and paragraphs to a 384-dimensional dense vector space.

The nli-mpnet-base-v2 model offers enhanced accuracy on semantic benchmarks, making it suitable for tasks requiring high semantic precision. This model takes a list of text snippets and returns a list of embeddings, where each embedding is a list of 768 floating-point numbers representing the semantic text embedding of the associated snippet. Text snippets can be up to 384 tokens in length (approximately 280 words).

Lastly, nv-embed-v2 is recognized as a top-performing generalist embedding model on the Massive Text Embedding Benchmark (MTEB), optimized for deployment on GPU infrastructure. These models differ in size and complexity, with transformer-based models such as MiniLM and MPNet designed to scale for superior semantic accuracy, while NVIDIA's embeddings are tailored for optimal performance on GPU platforms. nv-embed-v2 maps a text snippet 4096-dimensional dense vector.

## III. DOCUMENT SNIPPET REVIEW IN CONSTRUCTION DELAY LITIGATION

In the high-stakes environment of delay litigation, the ability to uncover and organize the facts surrounding project delays quickly is critical. Parties in delay litigations have strict time constraints around petitioning for or responding to delay related claims. Snippet review, which isolates and evaluates specific excerpts from documents, is an invaluable tool in identifying the potential causes of delays and constructing detailed chronologies of events. This section explores how snippet review supports the fact-finding process in delay litigation by improving precision, efficiency, and collaboration.

*A. Precision in Identifying Causes of Delay*

Delay litigation demands a meticulous examination of evidence to uncover all potential causes of delays on a project. Snippet review narrows the focus to the most relevant portions of documents, allowing litigation teams to pinpoint critical information such as contract obligations, communication records, and project milestones. By isolating these excerpts, snippet review ensures that even nuanced or indirect causes of delay are captured, laying a strong foundation for constructing accurate chronologies.

*B. Constructing Detailed Event Chronologies*

One of the primary objectives of delay reviews is to develop a comprehensive chronology of events leading up to the delays. Snippet review accelerates this process by extracting and organizing relevant text from extensive datasets. Delay related snippets provide clear, concise timeline of key events, enabling litigation teams to present a coherent narrative that aligns with the evidence. This structured approach strengthens the arguments and clarity of litigation submissions.

*C. Time and Cost Efficiency*

The volume of data involved in delay litigations can be daunting, requiring significant resources to review and analyze. Snippet review streamlines this process by automating the identification of pertinent information, thereby reducing the time and cost traditionally associated with manual document review. By optimizing resource allocation, legal and technical teams can focus their efforts on analysis and strategy, rather than data discovery.

*D. Enhanced Collaboration and Communication*

Litigation proceedings often involve multiple parties, including legal counsel, expert witnesses, and clients. Snippet review facilitates collaboration by delivering focused, contextually relevant excerpts that are easier for all parties to interpret. These concise snippets allow for more effective discussions and alignments on key issues, ensuring that the arbitration team operates with a unified understanding of the facts.

*E. Supporting Data-Driven Insights*

When combined with advanced analytics, snippet review offers powerful insights into project performance and systemic issues. By categorizing and analyzing snippets, litigation teams can identify patterns of delays, recurring inefficiencies, and potential breaches of contract. These data-driven insights provide a strategic advantage, enabling teams to present compelling, evidence-based arguments during arbitration proceedings.

In delay litigations, where uncovering the causes of delays and constructing accurate event chronologies are critical, snippet review provides a powerful solution. By enhancing precision, improving efficiency, and fostering collaboration, snippet review equips arbitration teams with the tools they need to navigate the complexities of delay disputes and deliver compelling, evidence-backed narratives.

## IV. EXPERIMENTS

### A. Dataset

The dataset used for this study consists of text snippets extracted from two concluded real-world construction delay dispute cases. These cases involve various document types, including emails, Microsoft Office files, PDFs, Excel spreadsheets, and other text-based formats. The documents cover a range of topics, with some related to construction delays and others not.

A total of 35,543 text snippets were extracted and meticulously classified by human reviewers as either construction delay-related or non-delay-related. Of these snippets:

• 4,357 snippets were identified as construction delay-related.

• 31,186 snippets were classified as non-delay-related.

The dataset is divided across two dispute matters, originating from different organizations:

1) The first dispute matter contains 3,240 snippets, including 762 delay-related snippets and 2,478 non-delay-related snippets. We refer this dataset as to dataset A.

2) The second dispute matter consists of 32,303 snippets, of which 3,595 are delay-related and 28,708 are not delay-related. We refer this dataset as to dataset B.

The combined dataset A and B is referred as to dataset AB. The dataset AB includes text snippets from two different matters. These datasets provide a diverse and realistic representation of real-world construction delay dispute scenarios, making it suitable for evaluating text classification models and comparing text embedding techniques in a practical context.

### B. Evaluation

To assess the performance of embedding models in the context of binary classification tasks, we used two supervised machine learning algorithms, K-Nearest Neighbor (KNN) and Logistic Regression (LR) to perform binary classification tasks on the three datasets. For the combined dataset AB, we used three-fold cross-validation to evaluate classification performance. We call this experiment as in-matter experiment. In a three-fold cross validation, all text snippets are randomly divided into three blocks and documents in one block were used as test documents, while the documents in the other two blocks were used as training documents. This process repeated three times, one for each block as test block. For dataset A and B, we conducted cross matter evaluation. We trained a text classification model using the documents in dataset A and tested the model on all documents in dataset B. Similarly, we trained a model using all documents in dataset B and tested the model on all documents in dataset A. These two experiments are called as cross matter experiments.

Two popular performance metrics: precision and recall, were used to measure binary classification performances. Recall is the ratio between the number of correctly identified delay snippets and the total number of delay snippets, whereas precision is the ratio between the number of correctly identified delay snippets and the number of snippets identified as delay snippets. We displayed classification performances with precision-recall curves.

In all experiments, all text snippets were first converted into numeric embeddings using the bag-of-word method and three different text embedding models discussed in Section II. Four models for each of the two learning algorithms were trained using the four different numeric embeddings respectively and tested on the corresponding embeddings of test documents. All embedding computations and model training processes were executed on Azure Cloud GPU server.

### C. Results

Figure 1 presents the precision-recall curves for KNN models (with K = 3) utilizing four different embedding methods on the combined dataset AB. These curves represent the average results from a three-fold cross-validation process. The three embedding methods demonstrated comparable performance, all of which surpassed the bag-of-words model. This outcome aligns with expectations, as embeddings are capable of capturing semantic similarities between snippets, whereas the bag-of-words model is limited to capturing syntactic similarities. The performance of KNN is significantly influenced by the similarity between snippets. The three embedding models achieved similar results.

Figure 2 displays the precision-recall curves for Logistic Regression using four different embedding methods on the combined dataset AB. These curves represent the average results from a three-fold cross-validation. Surprisingly, the bag-of-words model for logistic regression performed on par with the Nvidia model and outperformed the other two embedding models. This indicates that when there are sufficient high-quality training examples, logistic regression can effectively utilize the bag-of-words representation. MiniLM-L6-v2 obtained better results than nli-mpnet-base-v2.

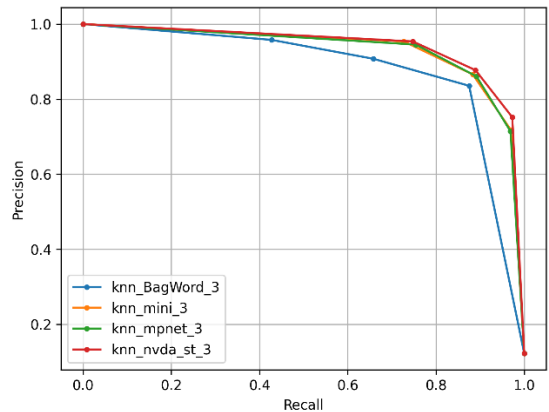

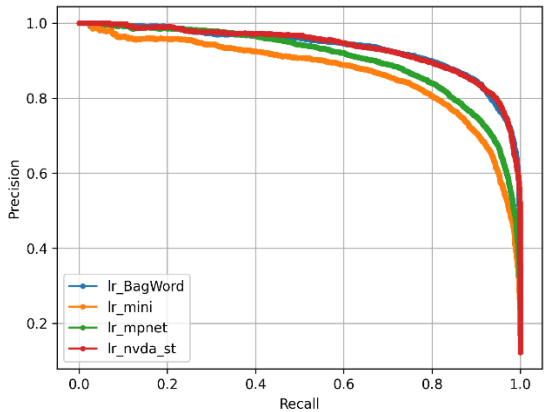

Fig. 1. Precision and Recall Curves for KNN with $k = 3$ Using Different Embedding Models for the combined dataset AB

Fig. 2. Precision and Recall Curves for LR Using Different Embedding Models for the combined dataset AB

Figures 3 and 4 present the precision-recall curves for KNN (K = 3) applied to Datasets A and B in cross-matter tests, respectively. In these tests, one dataset was used for training while the other served as the test dataset. It is evident that the bag-of-words representation performed significantly worse than the three embedding models. This is expected, as delay snippets may not be syntactically similar across matters, but they do share semantic similarities. Among the three embedding models, the Nvidia model outperformed the others, with MiniLM-L6-v2 surpassing nli-mpnet-base-v2.

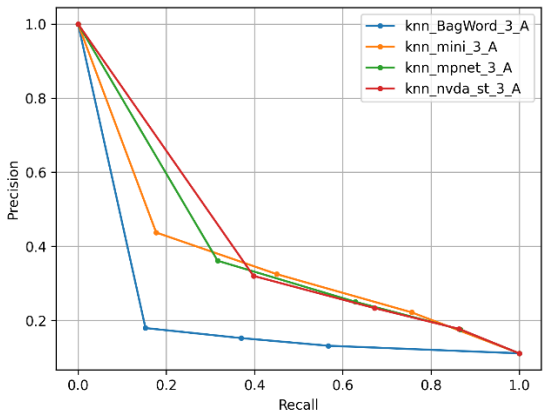

Fig. 3. Precision and Recall Curves for KNN with $k = 3$ Using Different Embedding Models for Dataset A as training set

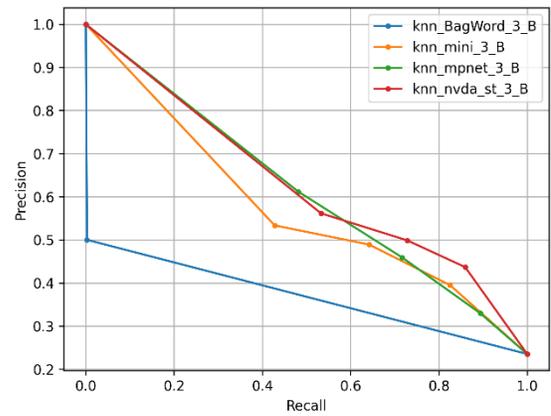

Fig. 4. Precision and Recall Curves for KNN with $k = 3$ Using Different Embedding Models for Dataset B as training set

Figures 5 and 6 illustrate the precision-recall curves for logistic regression applied to Datasets A and B in cross-matter tests, respectively. In Dataset A, the bag-of-words representation performed significantly worse than the three embedding models. This is likely because the terms used in the delay snippets of Dataset A are not highly similar to those in Dataset B. Once again, the Nvidia model outperformed the other two embedding models, with MiniLM-L6-v2 surpassing nli-mpnet-base-v2.

For Dataset B, the performance differences among the four models are not significant for recall values greater than 20%. While the Nvidia model performed slightly better overall, the nli-mpnet-base-v2 model struggled when recall values were below 20%.

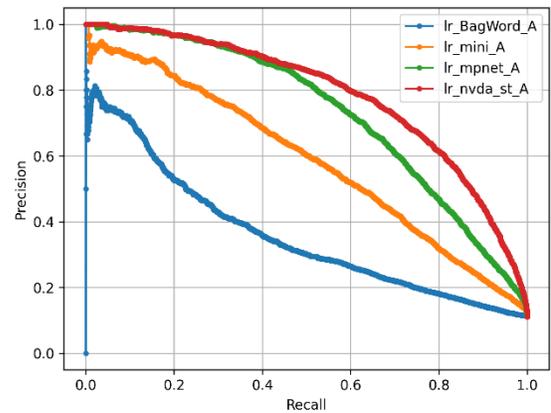

Fig. 5. Precision and Recall Curves for LR Using Different Embedding Models for Dataset A as training dataset

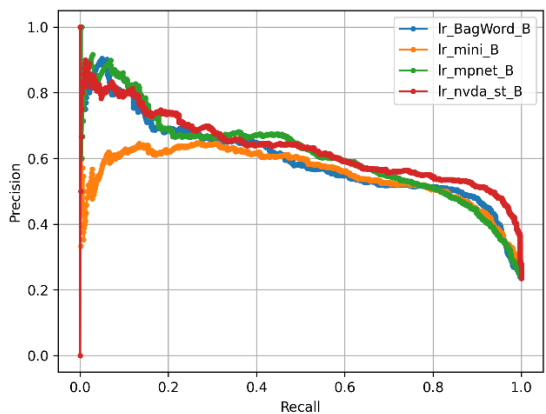

Fig. 6. Precision and Recall Curves for LR Using Different Embedding Models for Dataset B as training dataset

From our experiments, the NVIDIA embedding model outperformed all others consistently across all experiments, including both cross-validation and cross-matter validation settings. Embedding models show advantages over the bag-of-word model when training snippets are not syntactically similar to the test documents. Logistic regression can perform well when sufficient high quality training documents are available. The computational cost for embedding models is a significant drawback. An embedding model requires much longer processing times than the bag-of-word model. It is interesting to observe that the bag-of-word model performed almost as well as the NVIDIA model in the cross-validation setting when used with logistic regression. In contrast, the bag-of-word model exhibited poor performance in cross validation with KNN.

## V. FURTHER WORK

Building on the current evaluation of embedding models through binary classification tasks, future research will focus on expanding the scope of performance measurement to other key language tasks, such as information retrieval and question answering. Specifically, we are exploring the application of Retrieval-Augmented Generation (RAG) for information retrieval, comparing the four embedding models in handling diverse query-response scenarios. This will help assess the adaptability of embeddings to real-world tasks requiring semantic understanding and contextual relevance. Furthermore, we plan to extend the evaluation framework by integrating business-specific use cases, enabling us to analyze the practical impact of embedding models in domain-specific applications. These efforts aim to provide a comprehensive understanding of embedding model performance across diverse scenarios, facilitating better model selection for specialized tasks.

## REFERENCES

[1] Reimers, N., & Gurevych, I. (2019). Sentence-BERT: Sentence Embeddings using Siamese BERT-Networks. arXiv. Retrieved from https://arxiv.org/abs/1908.10084
[2] NVIDIA NeMo Documentation. NVIDIA AI. Retrieved from https://developer.nvidia.com/nemo.
[3] Raffel, C., et al. (2020). Exploring the Limits of Transfer Learning with a Unified Text-to-Text Transformer. Journal of Machine Learning Research. Retrieved from https://jmlr.org/papers/v21/20-074.html.
[4] Reimers, N., & Gurevych, I. (2019). Sentence-BERT: Sentence Embeddings using Siamese BERT-Networks. Proceedings of the 2019 Conference on Empirical Methods in Natural Language Processing, pp. 3982–3992. Association for Computational Linguistics. Available at: https://arxiv.org/abs/1908.10084.
[5] NVIDIA Corporation. (2023). NVIDIA AI Model Catalog. NVIDIA Developer. Available at: https://developer.nvidia.com/ai-catalog.